\documentclass{ecai}  
\usepackage{tikz}

\usepackage{graphicx}
\usepackage{latexsym}

\usepackage{hyperref}
\usepackage{subcaption}
\usepackage{amsmath}
\usepackage{amsthm}
\usepackage{booktabs}
\usepackage{algorithm}
\usepackage{algorithmic}
\usepackage{multirow}

\usepackage{pdfpages} % include pdfs
\usepackage{pgffor} % for loops

\makeatletter
\AtBeginDocument{\let\LS@rot\@undefined}
\makeatother

\def\supplementfilename{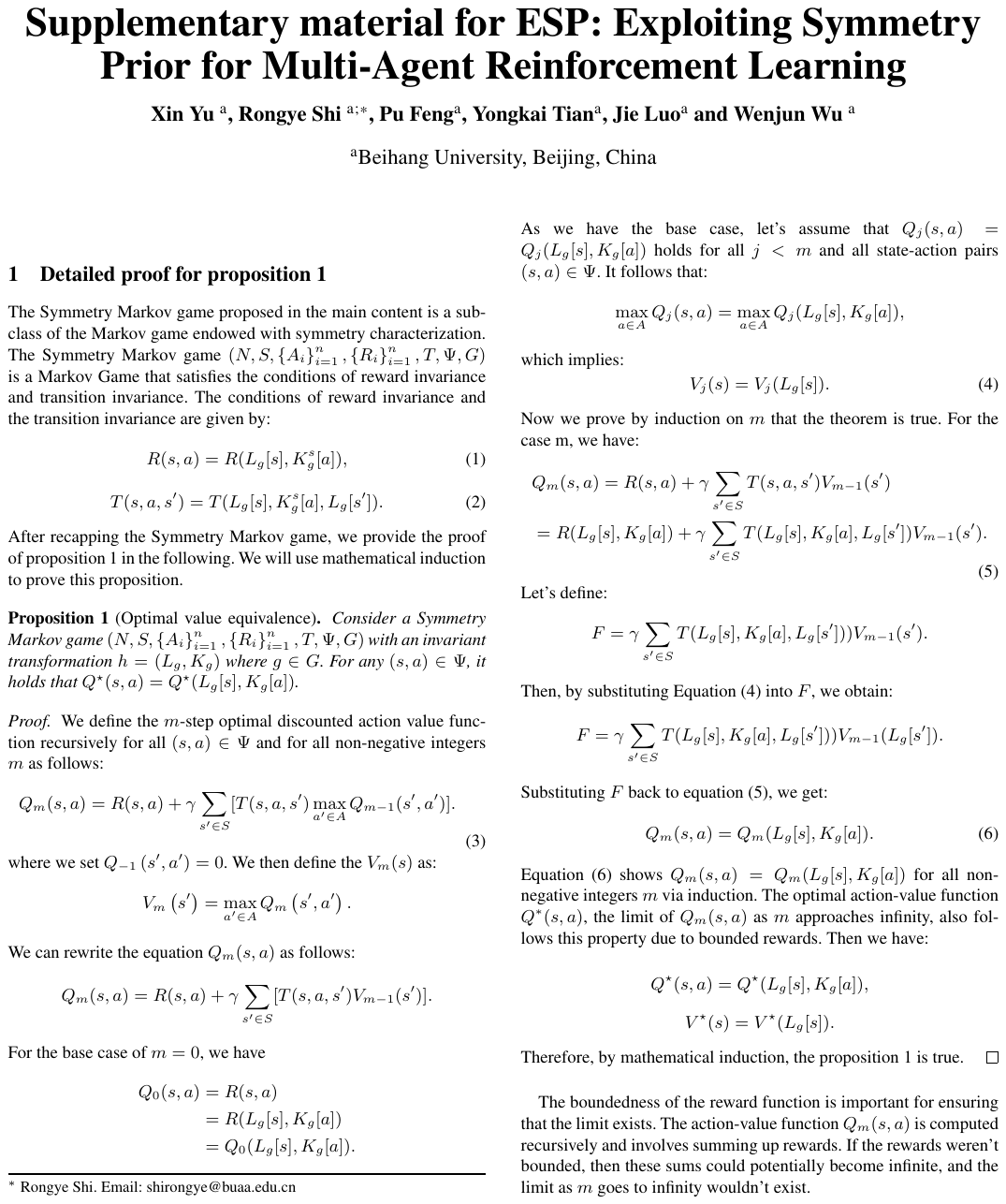}

\pdfximage{\supplementfilename}
\def\numbersupplementpages{\the\pdflastximagepages}

\newif\ifarXiv
\arXivtrue 
\begin{document}

\begin{frontmatter}

\title{\vspace{-2cm} {\scriptsize  \textcolor{blue}{This paper is accepted by the 26th European Conference on Artificial Intelligence (ECAI2023). Please cite it as:} \\  \textcolor{blue}{Xin Yu, Rongye Shi, Pu Feng, Yongkai Tian, Jie Luo, and Wenjun Wu. ESP: Exploiting Symmetry Prior for Multi-Agent Reinforcement Learning. \textit{The 26th European}} \\ \vspace{-0.5cm} \textcolor{blue}{ \textit{Conference on Artificial Intelligence (ECAI)}, 2023.}} \\ ESP: Exploiting Symmetry Prior for Multi-Agent Reinforcement Learning}

\author[A]{\fnms{Xin}~\snm{Yu} }
\author[A]{\fnms{Rongye}~\snm{Shi} \thanks{Rongye Shi is the corresponding author. Email: shirongye@buaa.edu.cn.}}
\author[A]{\fnms{Pu}~\snm{Feng}}  
\author[A]{\fnms{Yongkai}~\snm{Tian}} % use of \orcid{} is optional
\author[A]{\fnms{Jie}~\snm{Luo}} % use of \orcid{} is optional
\author[A]{\fnms{Wenjun}~\snm{Wu} } % use of \orcid{} is optional

\address[A]{Beihang University, Beijing, China}

\newtheorem{proposition}{Proposition}

\begin{abstract}
Multi-agent reinforcement learning (MARL) has achieved promising results in recent years. However, most existing reinforcement learning methods require a large amount of data for model training. In addition, data-efficient reinforcement learning requires the construction of strong inductive biases, which are ignored in the current MARL approaches. Inspired by the symmetry phenomenon in multi-agent systems, this paper proposes a framework for exploiting prior knowledge by integrating data augmentation and a well-designed consistency loss into the existing MARL methods. In addition, the proposed framework is model-agnostic and can be applied to most of the current MARL algorithms. Experimental tests on multiple challenging tasks demonstrate the effectiveness of the proposed framework. Moreover, the proposed framework is applied to a physical multi-robot testbed to show its superiority. 
\end{abstract}

\end{frontmatter}
\section{Introduction}

Artificial intelligence (AI) has been extensively applied across diverse scientific disciplines~\cite{shi2020improving,shi2021tits}. A variety of AI challenges can be formulated as multi-agent reinforcement learning (MARL) problems~\cite{feng2023mact}. With recent advances in MARL, many achievements in constructing intelligent agents have been realized, thus overcoming complex challenges, such as biology~\cite{yu2021swarm}, multiplayer games~\cite{dota2} and multi-robot tasks~\cite{ijrr}. However, the majority of the reinforcement learning (RL) methods strongly rely on massive amounts of data for model development. From a practical perspective, although a parallel and accelerated simulation environment enables agents to solve complex tasks within a reasonable amount of time, learning in real-world applications suffers from physical conditions-related constraints. Similarly, in simulations, due to limitations on the rendering speed, data efficiency is critical for realizing rapid experimental iterations~\cite{airsim,isacgym}. Therefore, improving the sample efficiency of the existing MARL methods is essential for both theoretical and practical research.

\begin{figure}[ht]
\centerline{\includegraphics[width=7cm]{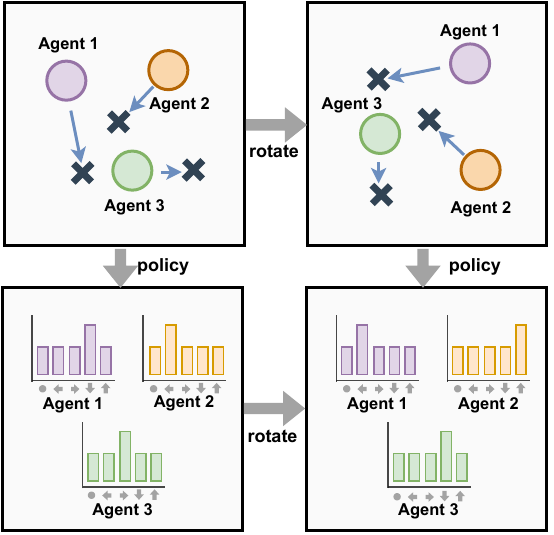}}
\caption{Illustration of multi-agent symmetry. Rotating the state globally in the cooperative navigation task results in a permutation of the optimal joint policy.}
\label{example}
\end{figure}

Improving sample efficiency entails reducing interactions with the environment in order to learn a specific policy more effectively. From the perspective of representation learning, achieving data-efficient reinforcement learning requires constructing stronger inductive biases, which poses a challenging task~\cite{pinnsurvey}. Recent literature has proposed various methods to address the data inefficiency issue in deep RL, broadly categorized as the data augmentation approach or network structure design approach. Specifically, data augmentation has been widely used in recent studies in single-agent RL, most of the proposed solutions focus on image-based data augmentation~\cite{visionaug}. A technique called RL with augmented data (RAD) incorporates multiple data augmentations into visual observations, such as image crop, enabling single-agent RL pipelines to outperform prior state-of-the-art methods~\cite{rlaug}. Despite the importance of data augmentation, to the best of our knowledge, theoretically founded data augmentation methods in MARL have been scarcely investigated. Second, network structure design approaches usually aim to design specialized network architectures that implicitly embed prior knowledge of a given task. Network structure design methods use specific sets of basis functions chosen according to the type of transformation group~\cite{eqcnn,stcnn}. For example, circular harmonic functions and Hermite polynomials have been used to achieve rotation and scale equivariance, respectively~\cite{stee, scale}. However, previous studies have shown that this type of constraint can overly restrict the model and result in sub-optimal performance on learning tasks~\cite{subopt}. In addition, the sets of basis functions vary for different transformation groups, making it challenging to design composite networks that are equivariant to multiple transformation groups such as rotations, reflections, and changes in scale. Compared to network structure design methods, our method enables the simultaneous utilization of multiple forms of symmetry, providing a more straightforward implementation. 

The most general symmetry in the multi-agent systems is the global symmetry shown in Figure \ref{example}, where rotating the global state results in a permutation of the optimal joint policy. Motivated by the symmetric phenomenon in multi-agent domains, we propose a general framework for MARL to increase sample efficiency. The proposed framework consists of two parts, a symmetry augmentation module, and a well-designed symmetry consistency loss. The symmetry augmentation module performs transformations on the trajectories to generate extra data before storing them in a replay buffer. Moreover, a symmetry consistency loss, which serves as an auxiliary module, is introduced to bring symmetry prior into an RL agent.

We conduct extensive experiments on several challenging MARL benchmarks (e.g., Predator-Prey, Cooperative Navigation, and Formation Change) to demonstrate the superior performance of the proposed framework regarding data efficiency and higher evaluation rewards. Experimental results demonstrate that our proposed method improves the performance of MARL on both simple and complex tasks. Compared to network structure design methods, our approach offers superior performance and is more straightforward to implement. In addition, we tested the performance of our methods using physical robots on the Formation Change task. Video demonstrations and Supplementary material are available at the project website \url{https://xinyu-site.github.io/esp-marl/}.
\section{Related work}

\subsection{Data augmentation in RL}

Introducing data augmentation into RL has primarily aimed at enhancing data efficiency~\cite{otherplay}. A natural approach to exploit data augmentation in single-agent RL is to obtain more data via image transformation during model training~\cite{rlaug,imageallneed,ITER,arm}. Another type of approach introduces data augmentation through an innovative contrastive learning framework called the CURL. The CURL learns representations that can improve data efficiency by enforcing consistencies between an image and its augmented version through instance contrastive loss~\cite{curl}. Although improving data efficiency is a widely studied problem in the RL field, fewer studies focused on exploiting the data augmentation methods in multi-agent domains. To the best of the authors’ knowledge, the most relevant algorithm from the available literature to this work is the data augmentation method for MARL proposed in~\cite{homoaug}. This method generates extra data by performing permutation transformations for homogeneous agents. The approach of performing data augmentation using permutation invariance is essentially another interpretation perspective of the parameter-sharing learning paradigm. However, this study goes beyond permutation symmetries to a broader group of symmetries in multi-agent RL.

\subsection{Integrating inductive bias into network}

Currently, the major approach for exploiting inductive bias in neural networks is to design specialized architectures that implicitly embed prior knowledge associated with a given task~\cite{geometric,report,symo,shi2021physics}. In the context of multi-agent reinforcement learning (MARL), graph neural networks (GNNs) are commonly used to model interactions between agents and allow for a graph-based representation~\cite{jiangjiechuan,GRAPHABS}. However, more prior knowledge needs to be integrated into MARL beyond the graph structure. The content relevant to this work focuses on embedding symmetry into network. For single-agent RL, symmetries in the joint state-action space can be expressed in MDP homomorphic networks~\cite{homon}. $\mathrm{SO}(2)$-Equivariant RL enforce symmetry within the structure of their convolutional layers~\cite{so2}. As for MARL, Multi-Agent MDP Homomorphic Networks integrate the symmetries into a neural network, thus improving data efficiency~\cite{vand}. These methods essentially rely on the construction of filters from linear combinations of certain sets of basis functions. Previous studies have shown that constraining the weight parameterization space to specific basis sets overly constrains the model, and although this strategy makes the model equivariant to the desired transformation group, the performance on the actual learning task may be sub-optimal~\cite{subopt}. Furthermore, the method of network structure embedding requires redesigning standard networks by incorporating sets of complex basis functions that tend to vary for different transformation groups. This makes it difficult to design composite networks that are equivariant to multiple transformation groups, such as rotations, reflections, and changes in scale~\cite{stee,scale}. In contrast, our method enables the simultaneous utilization of multiple forms of symmetry and avoids the limitations of network structure embedding in terms of complexity and design.

%Our approach is straightforward yet effective and can be conveniently applied to a wide range of multi-agent tasks.

\section{Background}

\subsection{Group and transformations}
We provide a brief introduction to the concepts of groups and transformations in this section~\cite{bros}. A group $G$ is a set with a binary operator that have specific mathematical properties: identity, inverse, closure, and associativity. We will refer extensively to the group $\mathrm{SO}(2)$ and its cyclic subgroup $C_n$. $\mathrm{SO}(2)$ is the group of continuous rotations $\left\{\operatorname{R}_\theta: 0 \leq \theta<2 \pi\right\}$. $C_n$ is the discrete subgroup $C_n=\left\{\operatorname{R}_\theta: \theta \in\left\{\frac{2 \pi i}{n} \mid 0 \leq i<n\right\}\right\}$ of rotations.

A rotation matrix is a transformation matrix that describes performing a rotation in Euclidean space~\cite{rotationm}. For a set of rotations $\left\{0^{\circ}, 90^{\circ}, 180^{\circ}, 270^{\circ}\right\}$, the rotation matrix is defined as follows:
\begin{equation}
    R(\theta)=\left[\begin{array}{cc}\cos \theta & -\sin \theta \\ \sin \theta & \cos \theta\end{array}\right]\label{rotation1}.
\end{equation}
The four group axioms are satisfied in the case of a rotation transformation. This paper primarily utilizes the $C_4$ group, while our method can be applied to other groups as well.
\subsection{Equivariance and invariance}

The symmetry appearing in multi-agent systems can be denoted as equivariance and invariance.
Given a transformation operator $L_g: \mathcal{X} \rightarrow \mathcal{X}$ and a mapping function $f: \mathcal{X} \rightarrow \mathcal{Y}$, if there exists a second transformation operator $K_g: \mathcal{Y} \rightarrow \mathcal{Y}$ in the output space of $f$ such that:
\begin{equation}
    K_g[f(x)]=f\left(L_g[x]\right),
\end{equation}
where $g \in G$ and $G$ is a mathematical group, then, function $f$ is equivariant to the transformation $L_g$. A related notion to equivariance is invariance. If for any choice of $g \in G$ we have that $K_g=I$, the identity function, then we say function $f$ is invariant to transformation $L_g$.

Figure \ref{example} shows the equivariance of the optimal policy, rotating the state globally results in a transformation of the optimal joint policy. Given two states $s$ and $L_g[s]$, the optimal policy $\pi^*$ is equivariant to the transformation $L_g$ which is denoted by $ K_g[\pi^*(s)]=\pi^*\left(L_g[s]\right)$. 

\subsection{Multi-agent proximal policy optimization}

Multi-agent Proximal Policy Optimization (MAPPO) is a variant of PPO which is specialized for multi-agent settings. In the default formulation of MAPPO under the Decentralized Partially Observable Markov Decision Process (Dec-POMDP) framework, the agents employ the trick of policy sharing. As such, the MAPPO learns policy $\pi_{\theta}$ by optimizing the following objective:
\begin{equation}
J_\pi(\theta)=\sum_{i=1}^n E_{ \pi_{\theta_{\text {old }}}}\left[\frac{\pi_\theta\left(a^i \mid s\right)}{\pi_{\theta_{\text {old }}}\left(a^i \mid s\right)} A^\pi\left(s, a^i\right)\right],\label{mappo_pi}
\end{equation}
where $\pi_{\theta_{\text {old }}}$ is the behavior policy used to collect trajectories, $\pi_\theta$ is the policy we want to optimize, and $A^\pi\left(s, a_t^i\right)$ denotes the advantage function. And the critic network is trained to optimize the following objective:
\begin{equation}
J_{V^\pi}(\psi)=-\sum_{i=1}^n E_{\pi}\left[V_{\text {target }}^\pi(s)-V_\psi^\pi(s)\right]^2.\label{mappo_v}
\end{equation}
In the above equations, $V_{\text {target }}^\pi(\boldsymbol{s})$ denotes the target value computed by Generalized Advantage Estimation~\cite{mappo}. For ease of clarity, we simplified the description of the objective. We use $s$ to denote the input of both policy and value functions, while the actual input may not be the same. 

\section{Problem statement}
\subsection{Cooperative markov game}
We formulate the cooperative MARL problem as cooperative Markov game~\cite{markov}. An $n$-agent cooperative Markov game can be represented using a tuple $(N,S,\left\{A_{i}\right\}_{i=1}^{n}, R, T,\Psi)$, where $N$ denotes the set of agents, $S$ is the state space, and $A_{i}$ is the action space of an agent $ i=1, \ldots, n $. Let $A=A_{1} \times A_{2} \times \cdots \times A_{n}$ be the joint action space. $T: S \times A \times S \rightarrow[0,1]$ is the transition function. $\Psi$ is the set of admissible state-action pairs. 
At time step t, the agents are at state $s_t$ (which may not be fully observable); the agents take independent action $(a_1,..., a_N)$ rely on their policy. Then, the environment emits the bounded joint reward $R$ and moves to the next state $s_{t+1}$.
The agents aim to maximize the expected joint return, defined as $E_{\pi}\left[\sum_{t=0}^{\infty} \gamma^t R\left(s_t, a_t\right)\right]$, where $\gamma$ is the discount factor, by selecting actions according to the policy $\pi_{i}:{S} \times{A}_{i} \rightarrow[0,1]$. The initial states are determined by a distribution $\eta:{S} \rightarrow[0,1]$.

\subsection{Symmetry markov game}
We propose Symmetry Markov game which is a subclass of the cooperative Markov game endowed with symmetry characterization. The Symmetry Markov game $(N, S,\left\{A_{i}\right\}_{i=1}^{n}, R, T,\Psi, G )$ is a cooperative Markov Game that satisfies the conditions of reward invariance and transition invariance. The state transformation and action transformation in the Symmetry Markov game are defined as $L_{g}: S \rightarrow S$ and $K_{g}: A \rightarrow A$, respectively. For state-action pairs $(s,a) \in \Psi$, we denote the transformation of group $g$ on $s$ and $a$ as $(L_{g}[s],K_{g}[a]) \in \Psi$. Further, for all $g \in G$, $s \in S$, $a \in A$, the conditions of reward invariance and the transition invariance are given by: 
\begin{equation} 
        R(s, a)=R(L_{g}[s], K_{g}[a]), \label{Rg} 
\end{equation} 
\begin{equation} 
     T(s, a, s^{\prime})=T(L_{g}[s],K_{g}[a],L_{g}[s^{\prime}]). \label{Tg}
\end{equation}
We assert that the symmetry property is a common occurrence in many real-world multi-agent tasks, albeit often requiring effort to identify. One approach to identifying symmetry, for instance, is to observe whether an optimal action remains optimal for a transformed state. If so, the symmetry property may be present. It should be emphasized that our method is not restricted to the Symmetry Markov game and can also be applied to Dec-POMDPs with symmetry. Additional details on this topic are available in the Sec.4 of the Supplementary Material.

\section{Methods}

\begin{figure*}[htbp]
\centerline{\includegraphics[width=15.5cm]{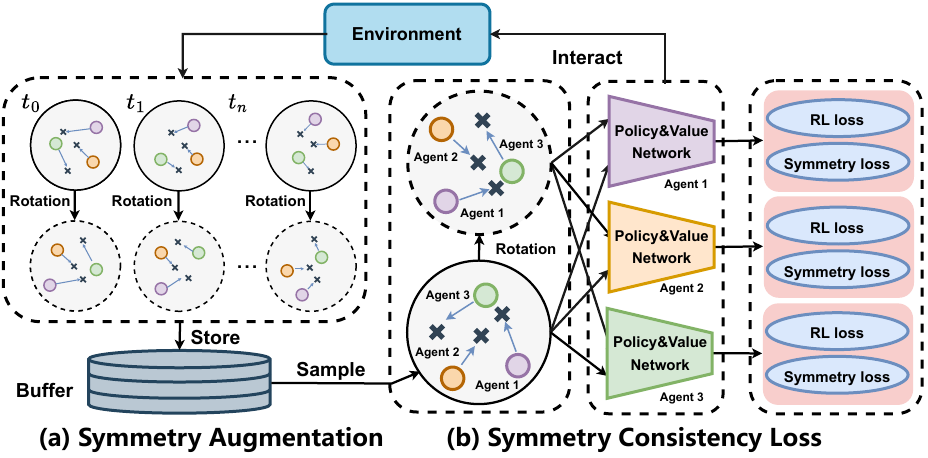}}
\caption{The overall framework of the proposed ESP. The ESP consists of the symmetry augmentation module and symmetry consistency loss module. symmetry augmentation module: the policy interacts with the environment to generate samples, which are then augmented and deposited into a buffer. symmetry consistency loss: transitions are sampled from the buffer, and the data are transformed to obtain symmetric transition pairs, which are subsequently used to compute symmetry consistency loss and further update the agent network.}
\label{framework}
\end{figure*}
%In this section, we present a general framework for exploiting symmetry prior, referred to as ESP.

In this section, we introduce our proposed framework, the Symmetry Prior Exploitation (ESP), for improving the performance of existing MARL algorithms. It is important to note that although fully utilizing symmetry is a desirable property in certain deep learning applications, our primary objective is to maximize the performance of agents in reinforcement learning tasks. To achieve this goal, we avoid using a hard constraint approach that embeds symmetry into the network structure. Instead, we treat symmetry as an additional objective and incorporate it through soft constraints such as data augmentation and regularization. This allows us to improve the performance of the reinforcement learning algorithm while benefiting from the advantages of symmetry. ESP comprises of two main components: a symmetry augmentation module and a symmetry consistency loss module.

\subsection{Symmetry augmentation} 
The most straightforward way to utilize symmetry is through data augmentation. This approach has been widely used in deep learning and has been shown to be effective in various tasks. In the context of multi-agent reinforcement learning, data augmentation can be used to generate additional samples that reflect the symmetry property of the environment, which can improve the training efficiency of the agents. Although data augmentation has been widely used in supervised learning, it still lacks a theoretical framework to prove the correctness of exploiting symmetry to augment state-action pairs in MARL. This study proposes a multi-agent optimal value function equivalence proposition for the symmetric Markov game. 

In a Symmetry Markov game, the transition and reward functions are invariant to group elements $g \in G$ acting on the state and action space. In the following, variables without a subscript $i$ denote the concatenation of all variables for all agents. (e.g. $a$ denotes the joint actions of all agents).

\begin{proposition}[Optimal value equivalence] \label{prop1}
Consider a Symmetry Markov game $(N,S,\left\{A_{i}\right\}_{i=1}^{n},\left\{R_{i}\right\}_{i=1}^{n}, T,\Psi,G)$ with a invariant transformation $h=(L_{g},K_{g})$ where $g \in G$. For any $(s, a) \in \Psi$, it holds that $Q^{\star}(s, a)=Q^{\star}(L_{g}[s],K_{g}[a])$.
\end{proposition}

A detailed proof of proposition \ref{prop1} is provided in Sec.1 of the Supplementary Material. Proposition \ref{prop1} indicates that using transformation data to train the policy in a Symmetry Markov game is reasonable. The process of utilizing data augmentation in MARL is demonstrated in Figure \ref{framework}, where the policy interacts with the environment to generate samples that are subsequently augmented. These trajectories can then be stored in the replay buffer and used for training purposes. This will be explained in the example of a multi-agent navigation task, which is shown in Figure \ref{example}. In this task, agents need to cover three target locations in a two-dimensional space simultaneously while avoiding collisions with each other. Agents observe the relative positions of other agents and landmarks. When performing a rotation operation on a state, each element can be multiplied by the rotation matrix given in Equation \eqref{rotation1}. After multiplying the action vector by the permutation matrix, a set of trajectories consisting of true trajectories and their symmetric trajectories is obtained. It is worth mentioning that the proposed symmetry augmentation method is not limited by the type of algorithm and can be directly applied to most MARL algorithms.

\subsection{Symmetry consistency loss}

While data augmentation has been previously utilized in RL, we argue that there are issues with its application in practice. For clarity of presentation, the MAPPO is used as an example to introduce the symmetry consistency loss. If transformation $(L_{g},K_{g})$ is applied to the algorithm, the MAPPO objective changes, and equation \eqref{mappo_pi} is replaced by

\begin{equation}
J_\pi(\theta)=\sum_{i=1}^n E_{\pi_{\theta_{\text {old }}}}\left[\frac{\pi_\theta\left(K_{g}[a^i] \mid L_{g}[s]\right)}{\pi_{\theta_{\text {old }}}\left(a^i \mid s\right)} A_\psi^\pi\left(s, a^i\right)\right].\label{map}
\end{equation}
However, the right hand is not a sound estimate of the left hand because $\pi_\theta(K_{g}[a] \mid L_{g}[s]) \neq \pi_\theta(a \mid s)$, particularly in the early stage of training. In fact a certain transformation $(L_{g},K_{g})$ for data augmentation can result in an arbitrarily large ratio $\pi_\theta(K_{g}[a] \mid L_{g}[s]) / \pi_{\theta_{o l d}}(a \mid s)$. Please see the Sec.2 of the Supplementary Material for more details. In multi-agent settings, using data augmentation to improve sample efficiency can be challenging. The reason for this is that when multiple agents are considered, more sources of variance are introduced, making the training severely unstable as shown in equation \eqref{map}. The proposed symmetry consistency loss provides a solution to this challenge and helps to make the use of data augmentation successful in multi-agent settings.

\textbf{Symmetry consistency loss}.
In order to properly utilize data augmentation in multi-agent reinforcement learning, we have designed a symmetry consistency loss for the policy and value function. The policy consistency loss term $S_\pi(\theta)$ is defined as
\begin{equation}
S_\pi=K L\left[\pi_\theta(K_{g}[a] \mid L_g[s]) \mid \pi_\theta(a \mid s)\right],\label{symloss1}
\end{equation}
aims to constrain distribution $\pi_\theta(K_{g}[a] \mid L_{g}[s])$ to be close to $\pi_\theta(a \mid s)$. This helps guide the training process according to the symmetry prior.

Assume that $V_\psi(s)$ represents an approximate value for state $s$, the symmetry consistency loss for value function is shown as
\begin{equation}
S_V=E_{s, a}\left[\left(V_\psi(s)-V_\psi\left(L_g[s]\right)\right)^2\right],\label{symloss2}
\end{equation}
designed to minimize the discrepancy between the outputs of the value function when provided with the original input and the symmetry-transformed input. Therefore, we regard Equation \eqref{symloss1} and \eqref{symloss2} as the symmetry consistency loss.

\textbf{MARL with symmetry consistency loss}.
It is convenient to combine the symmetry consistency loss with the existing MARL algorithms. For the policy-based algorithm MAPPO, our ESP framework maximizes the following objective:
\begin{equation}
    J_{ESP}=J_{MAPPO}-c (S_\pi+S_V),
\end{equation}
where $c$ is the hyperparameter that controls the symmetry consistency loss coefficient. $J_{MAPPO}$ are defined in Equation \eqref{mappo_pi} and \eqref{mappo_v}. As shown in Figure \ref{framework}, data sampled from the replay buffer is transformed to $(L_{g}[s], K_{g}[a])$, and the paired data are used to calculate the symmetry consistency loss. We can adjust hyperparameters $c$ according to the strength of symmetry to attempt to embed symmetry into the learning process.

\begin{figure*}[htbp]
     \centering
     \begin{subfigure}[b]{0.32\textwidth}
         \centering
         \includegraphics[width=\textwidth]{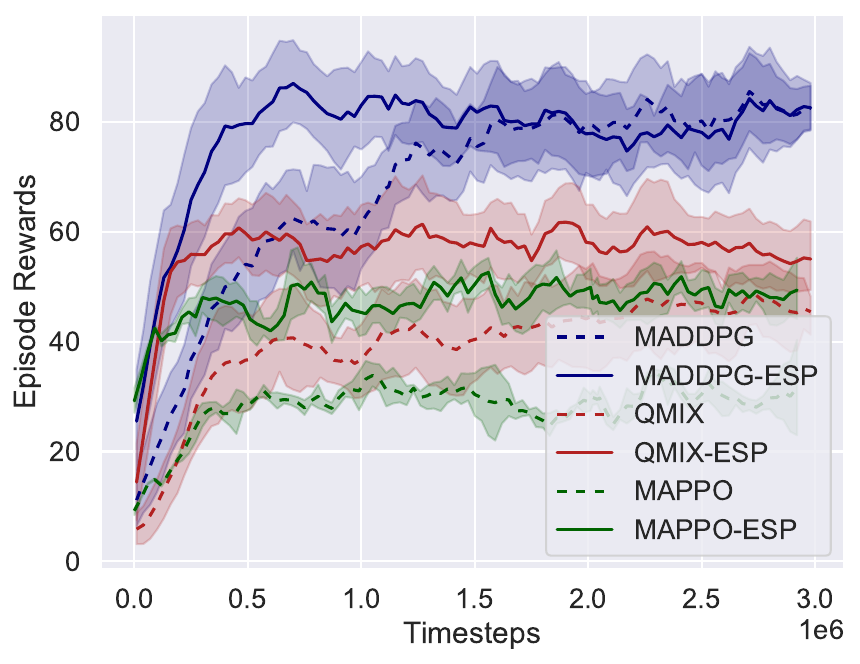}
         \caption{Predator-Prey}
         \label{4a}
     \end{subfigure}
     \hfill
     \begin{subfigure}[b]{0.32\textwidth}
         \centering
         \includegraphics[width=\textwidth]{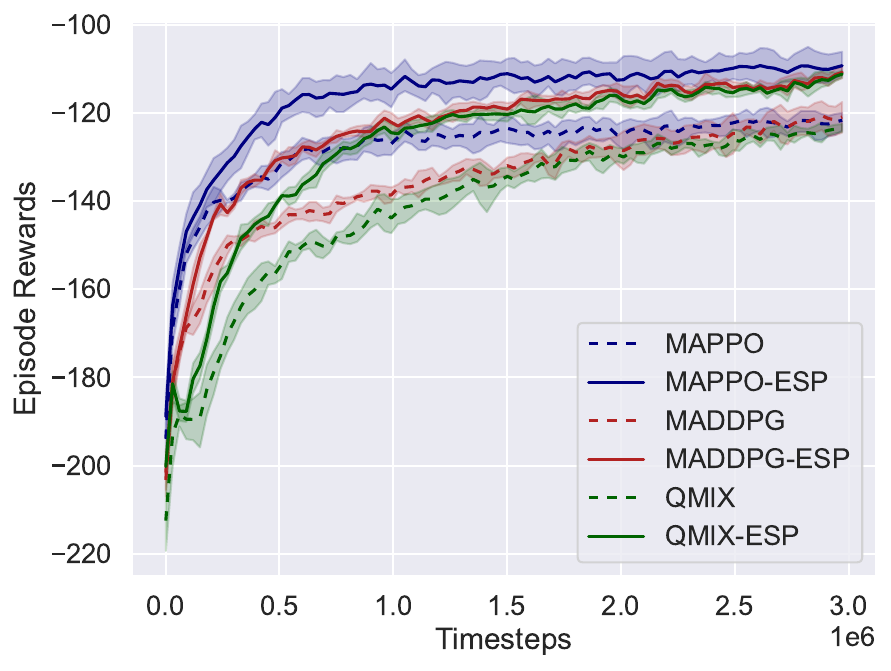}
         \caption{Cooperative Navigation}
         \label{4b}
     \end{subfigure}
     \hfill
     \begin{subfigure}[b]{0.32\textwidth}
         \centering
         \includegraphics[width=\textwidth]{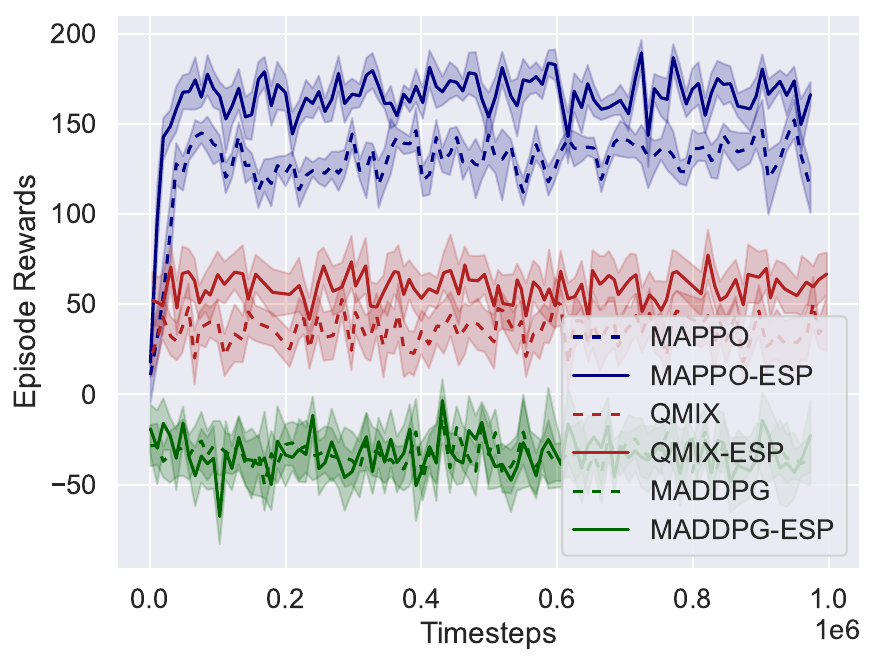}
         \caption{Formation Change}
         \label{4c}
     \end{subfigure}
        \caption{Learning curves of the original MADDPG, QMIX, and MAPPO and their versions with the ESP framework on the three multi-agent tasks. Each experiment was executed ten times with different random seeds. The solid lines denote the algorithm employing ESP framework.}
        \label{4}
\end{figure*}

\subsection{Algorithm description}
The training procedure of our ESP framework is described in Figure \ref{framework}. We start by initializing the group transformation $h=(L_{g},K_{g})$ alongside several training parameters, including transformation group $g$, agent numbers, maximum steps, and batch size. Then the transformation $h$ is selected according to group $g$. After that, the real trajectories generated by agents are augmented by symmetric transformation. Both real and augmented samples are stored in an experience replay buffer. Finally, data are sampled from the replay buffer to update the agent network using the symmetry consistency loss. Our method is based on data augmentation and is not limited by fully observable or partial observable. When facing partially observable problems, we can rotate the state globally during the centralized training phase, and then augment the agent's trajectories which can be used to train the algorithm. In the execution phase, the agents can perform the policy in a distributed manner.

\section{Experiments}

\subsection{Environmental setting}

The effectiveness of our framework was verified by experiments on four different tasks, cooperation navigation (CN), predator-prey (PP), formation change (FC), and wildlife monitoring. About the detailed settings of the task, such as reward function, state space, and action space, please refer to the Sec.5 of the Supplementary Material.

\begin{figure}[htbp]
     \centering
     \begin{subfigure}[b]{0.155\textwidth}
         \centering
         \includegraphics[width=\textwidth]{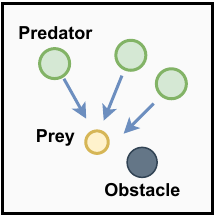}
         \caption{Predator-Prey}
         \label{schem1}
     \end{subfigure}
     \hfill
     \begin{subfigure}[b]{0.155\textwidth}
         \centering
         \includegraphics[width=\textwidth]{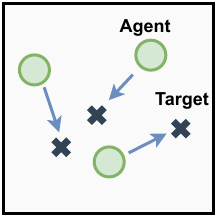}
         \caption{Navigation}
         \label{schem2}
     \end{subfigure}
     \hfill
     \begin{subfigure}[b]{0.155\textwidth}
         \centering
         \includegraphics[width=\textwidth]{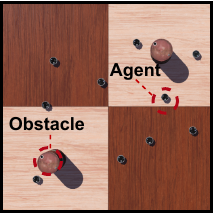}
         \caption{Formation Change}
         \label{schem3}
     \end{subfigure}
        \caption{The schematics of the experimental tasks.}
        \label{schem}
\end{figure}

\textbf{Predator-Prey}. Predator-Prey is a classic scenario implemented in a multi-agent particle environment~\cite{MPE}. In this task, collaborating agents need to catch prey. Predators receive a reward at every time step when at least one predator is on top of the prey. The Predator-Prey task is presented in Figure \ref{schem1}.

\textbf{Cooperative Navigation}. Cooperative navigation is a classic scenario implemented in a multi-agent particle environment. In this task, agents need to cover target landmarks whilst avoiding collisions with each other. Both agents and targets are spawned at random locations at the start of each episode. A classic cooperative navigation scenario with three agents is presented in Figure \ref{schem2}.

\textbf{Formation Change}. A multi-robot formation change task was designed inspired by~\cite{ijrr}. This task is shown in Figure \ref{schem3}, in this task, all robots started on a square formation and had their goals set to the opposite side. The E-puck was selected as a model in the open-source mobile robot simulator Webots~\cite{epuck}. The E-puck was controlled by assigning the wheel speed. Robots were required to learn to avoid obstacles and each other and coordinate to cover the destinations. Robots took actions according to the current state in each step and obtained the corresponding rewards, and then made the next state. When the maximum step number was reached, the current episode ended, and the next one started. 

\textbf{Wildlife monitoring.} The wildlife monitoring is a grid-world based environment, where a set of drones has to coordinate to accomplish the task~\cite{homon}. The goal is to trap poachers by having one drone hover above them while the other assists from the side. Two drones cannot be in the same location at the same time. This task was used to compare our ESP with the network design method. 

\textbf{Baselines}. The proposed framework was applied to several baselines, including Multi-Agent Deep Deterministic Policy Gradient (MADDPG), Monotonic Value Function Factorisation for Deep Multi-Agent Reinforcement Learning (QMIX), and Multi-Agent Proximal Policy Optimization (MAPPO), which are mainstream MARL approaches~\cite{qmix,maddpg,mappo}.

% iclr论文不能用在本文的三个任务上

\subsection{Main results}

\begin{table*}[htbp]
\caption{Episode rewards of the ESP and baselines on the three tasks. '500k' and '3000k' represent the number of training steps of the algorithms. The error bars are the standard error of the mean.}
\label{table1}
\centering
\begin{tabular}{cccccccc}
\toprule
Task & Steps & MADDPG & \textbf{MADDPG-ESP} & QMIX & \textbf{QMIX-ESP} & MAPPO & \textbf{MAPPO-ESP} \\
\midrule
\multirow{2}{*}{Predator-Prey} & 500k & 56.81$\pm$11.26 & \textbf{80.71$\pm$9.12} & 38.93$\pm$8.96 & \textbf{59.95$\pm$4.86} & 28.35$\pm$6.76 & \textbf{45.22$\pm$5.45} \\
 & 3000k & 85.43$\pm$8.14 & \textbf{87.06$\pm$4.87} & 48.78$\pm$7.75 & \textbf{57.75$\pm$6.81} & 32.54$\pm$8.21 & \textbf{59.54$\pm$4.32} \\
\multirow{2}{*}{Cooperative Navigation} & 500k & -145.23$\pm$3.71 & \textbf{-132.22$\pm$1.27} & -154.24$\pm$3.16 & \textbf{-140.12$\pm$2.25} & -132.65$\pm$2.89 & \textbf{-122.25$\pm$1.62} \\
 & 3000k & -122.63$\pm$6.04 & \textbf{-112.21$\pm$2.75} & -124.67$\pm$3.05 & \textbf{-113.84$\pm$1.43} & -126.97$\pm$2.12 & \textbf{-110.22$\pm$2.12} \\
\multirow{2}{*}{Formation Change} & 200k & -45.63$\pm$13.96 & \textbf{-41.13$\pm$11.17} & 48.53$\pm$20.12 & \textbf{51.23$\pm$14.71} & 122.56$\pm$15.12 & \textbf{163.18$\pm$14.14} \\
 & 1000k & -42.13$\pm$10.71 & \textbf{-44.13$\pm$12.31} & 49.44$\pm$18.71 & \textbf{59.47$\pm$14.35} & 140.64$\pm$18.45 & \textbf{170.85$\pm$10.2} \\
\bottomrule
\label{tableall}
\end{tabular}
\end{table*}

\begin{figure*}[htbp]
     \centering
     \begin{subfigure}[b]{0.32\textwidth}
         \centering
         \includegraphics[width=\textwidth]{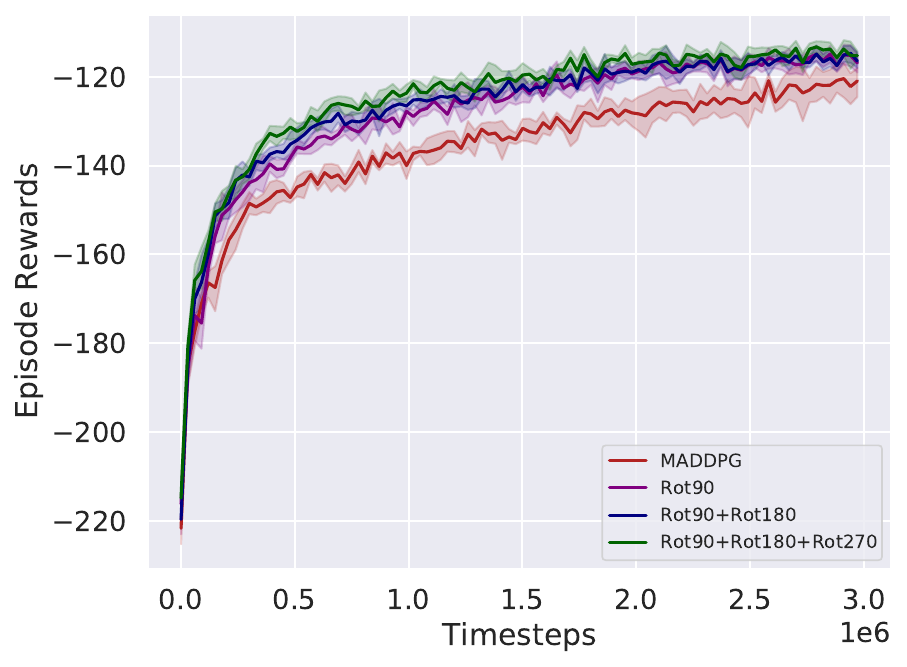}
         \caption{Learning curve of MADDPG using augmentation solely with different numbers.}\label{5a}
     \end{subfigure}
     \hfill
     \begin{subfigure}[b]{0.32\textwidth}
         \centering
         \includegraphics[width=\textwidth]{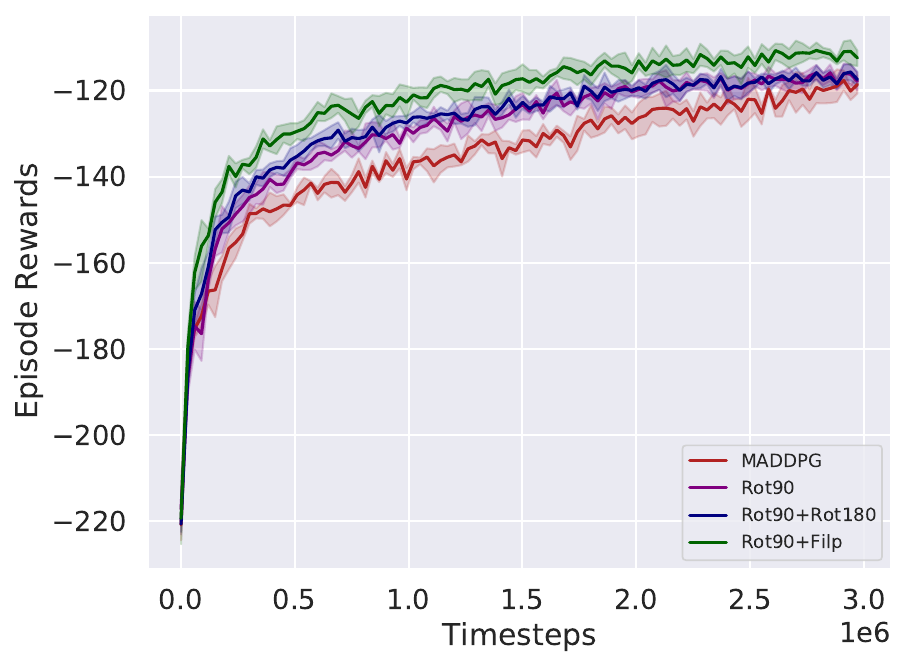}
         \caption{Learning curve of MADDPG using different augmentation types. }\label{5b}
     \end{subfigure}
     \hfill
     \begin{subfigure}[b]{0.32\textwidth}
         \centering
         \includegraphics[width=\textwidth]{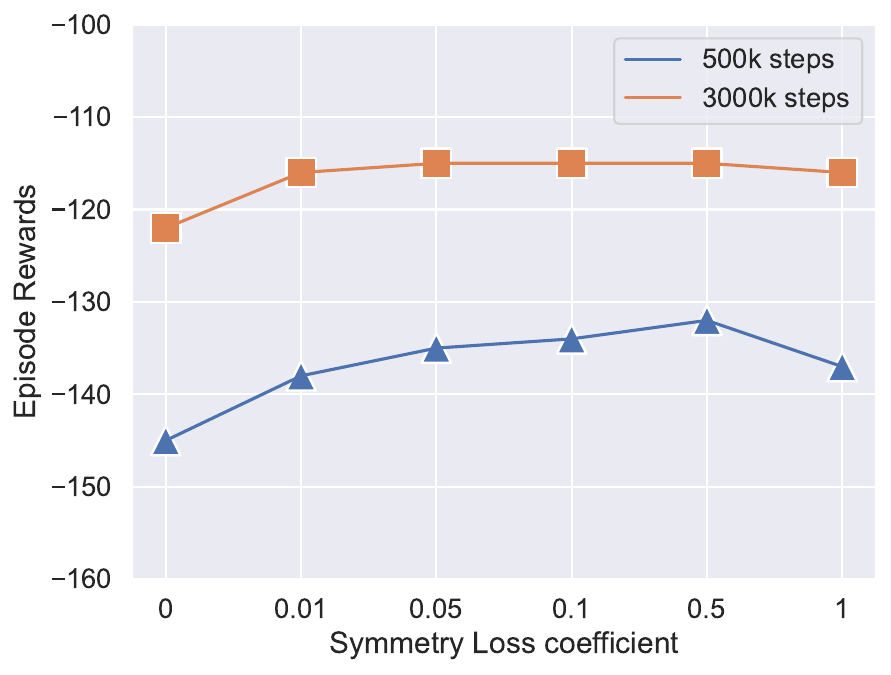}
         \caption{Parameter sensitivity of symmetry loss coefficient at different training steps.}\label{5c}
     \end{subfigure}
     \caption{The ablation study and parameter sensitivity of our ESP framework. }\label{5}
\end{figure*}

This section presents the experimental results obtained using the setup described in Section 6.1. The performance of each algorithm was evaluated with ten different random seeds, and the final experimental results are shown in Figure \ref{4} and Table \ref{table1}. The results show that the algorithms adopting the ESP framework achieved different degrees of advantage over their original versions.

\textbf{Predator-Prey.} In this scenario, there were three predators and one prey. As shown in Figure \ref{4a}, the proposed ESP framework outperformed the baseline methods significantly. The results indicated that the proposed framework could increase both the data efficiency and the convergence reward.

\textbf{Cooperative Navigation.} The cooperative navigation was a fully cooperative environment, where 3 agents (circles) cooperated to reach 3 landmarks (crosses) under a minimum number of collisions. Similarly, as shown in Figure \ref{4b}, the results show that the proposed framework can improve data efficiency and convergence rewards in this task.

\textbf{Formation Change.} To evaluate the proposed method in complex tasks, experiments were conducted on the multi-robot formation change task in the Webots simulator, as shown in Figure \ref{schem3}. In this scenario, eight robots started in a square formation and had their goals set on the opposite side. The experimental results showed that the MADDPG and QMIX could not learn a useful policy in this task, whereas the agents trained by the MAPPO-ESP and MAPPO could reach the goal while avoiding collisions with each other and obstacles. As presented in Figure \ref{4c}, the algorithms enhanced by the proposed framework obtained higher rewards than their original versions. This indicates that the proposed method can effectively improve the performance of multi-agent reinforcement learning algorithms in challenging environments.

In Table \ref{tableall}, the episode rewards of different algorithms at different training steps in the three task scenarios are presented. The superiority of the proposed framework is reflected in two main aspects. First, the proposed algorithm outperforms baseline algorithms by giving higher converged rewards. Additionally, it can be observed that the proposed algorithm exhibits faster convergence during training, possibly due to the utilization of prior knowledge within the framework. In contrast, the baseline trains the network through pure trial and error, which has a high computational cost. Consequently, the proposed method has strong versatility and can perform well in different tasks. 

\subsection{Ablation study}

To verify the effectiveness of each part in the proposed framework, ablation experiments were conducted in cooperative navigation task. Our experiments demonstrate that: 1) Using a single data augmentation significantly improves performance, whereas adding more of the same type has only a minor effect.  2) Using diverse transformation types is more effective than using multiple transformations of the same type. 3) The data augmentation and symmetry loss modules are both effective and improve performance when used in combination.

\textbf{Impact of the number and type of transformations}. In this section, we investigate the impact of using different numbers and types of symmetric transformations on performance without using symmetry loss. Firstly, we examine the impact of using different quantities of the same group of augmentations. As shown in Figure \ref{5a}, we used one to three rotation augmentations. We observed that the convergence speed of MADDPG gradually increased with an increasing number of rotations of the same type, and convergence reward slightly improved. Secondly, we analyze the impact of using different types of symmetry transformations. We denote the flip around the x-axis as Flip. As illustrated in Figure \ref{5b}, the simultaneous use of rotation and flip has a more pronounced effect on performance than using only rotation transformations.

\textbf{Effect of Symmetry Consistency Loss Coefficient}.
In this subsection, we analyze the impact of the symmetry consistency loss coefficient. We use the transformation of rotation by 90 degrees for the remaining experiments. The ESP was tested with different values of the symmetry consistency loss coefficient. In Figure \ref{5c}, the horizontal axis denotes the value of the symmetry consistency loss coefficient, and the vertical axis shows the testing episode reward averaged over ten seeds. The color of a line denotes a different step of the training steps. When the symmetry consistency loss coefficient equaled zero, the algorithm degraded to the baseline algorithm MADDPG. The results indicated that as the coefficient value increased, the episode rewards of the early training stage (i.e., 500k steps) first increased correspondingly and then decreased when the coefficient was larger than 0.5. However, the coefficient had little effect on the convergence episode rewards, at the training step of 3,000k step.

\begin{figure*}[htbp]
     \centering
     \begin{subfigure}[b]{0.32\textwidth}
         \centering
         \includegraphics[width=\textwidth]{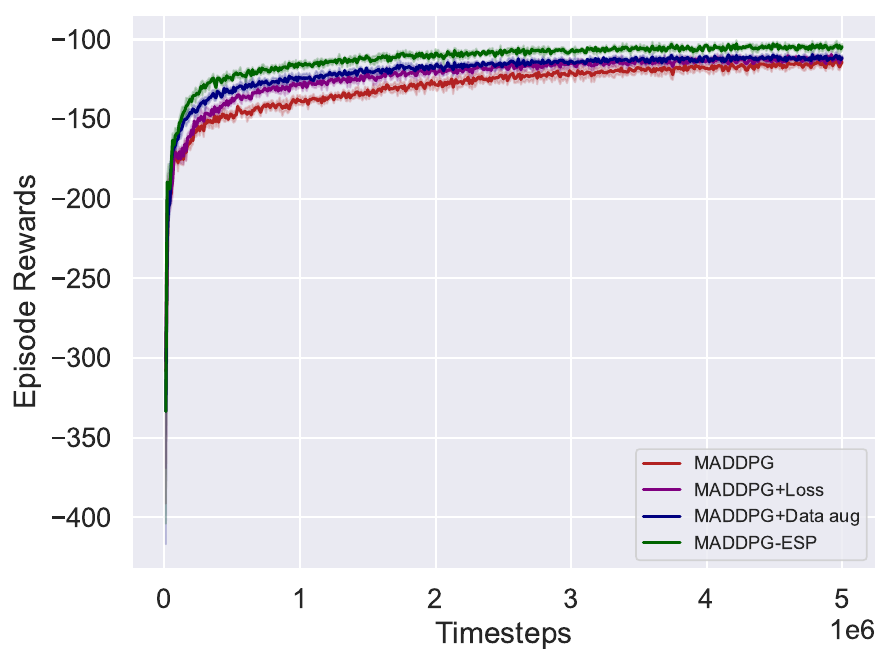}
         \caption{Learning curve of four variants of MADDPG in the task.}\label{6a}
     \end{subfigure}
     \hfill
     \begin{subfigure}[b]{0.32\textwidth}
         \centering
         \includegraphics[width=\textwidth]{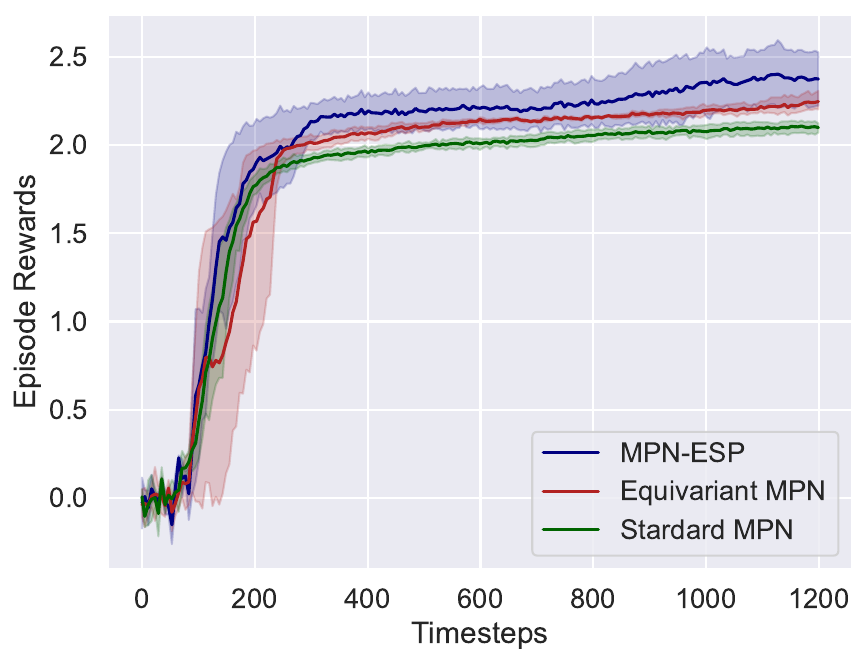}
         \caption{Results for the drone wildlife monitoring task shown over ten random seeds.}\label{6b}
     \end{subfigure}
     \hfill
     \begin{subfigure}[b]{0.32\textwidth}
         \centering
         \includegraphics[width=\textwidth]{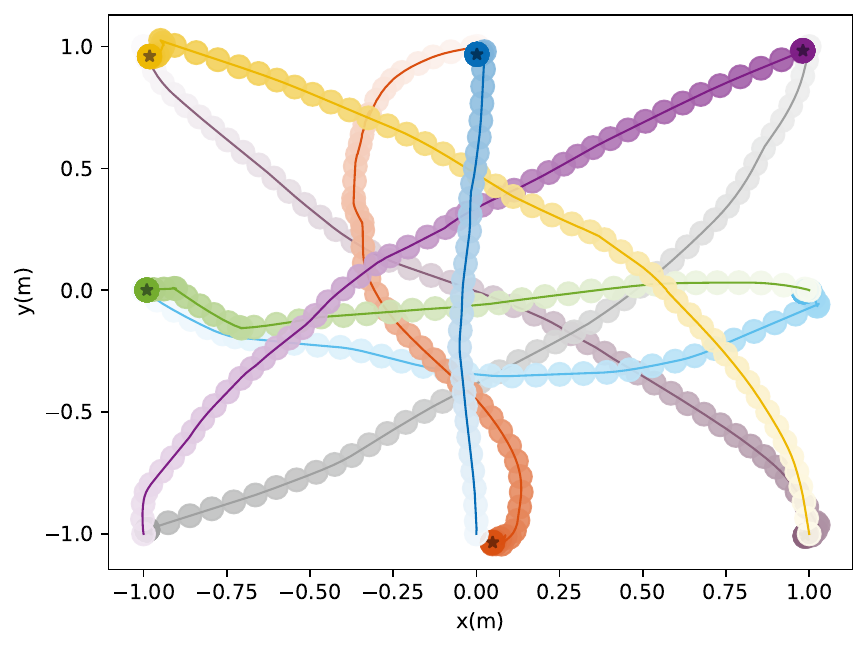}
         \caption{Trajectories of the robots executing the policy in the formation change task. }\label{6c}
     \end{subfigure}
     \caption{Additional experimental results for validating the effectiveness of the ESP framework. }\label{6}
\end{figure*}

\textbf{Effect of different modules}. The effect of data augmentation and symmetry consistency loss on the MADDPG performance was investigated in this subsection. We choose $c$= 0.5 to balance the traditional RL loss and the symmetry consistency loss. Four variations were analyzed:  1) MADDPG which is the vanilla algorithm; 2) MADDPG + loss, which used only the symmetry consistency loss in the MADDPG; 3) MADDPG + Data aug, which used only the symmetry augmentation in the training process; 4) MADDPG-ESP, which used the symmetry consistency loss and the symmetry augmentation in the MADDPG. As shown in Figure \ref{6a}, when used separately, both of the proposed modules improved the convergence speed and the final rewards. When used together, their combined performance was even better. The empirical results suggested the importance of the two modules in improving multi-agent task performance.

\subsection{Comparison with network design method}
In~\cite{vand}, the authors integrated symmetry into the MARL process by designing a specialized network structure, namely Multi-Agent MDP Homomorphic Networks, for wildlife monitoring tasks. However, Multi-Agent MDP Homomorphic Networks are limited to grid-world environments and rely on pixel-based states, making them challenging to extend to more complex scenarios. The three task scenarios shown in Figure \ref{schem} involve complex, continuous state spaces that cannot be handled by the network structure proposed in~\cite{vand}. Therefore, we compared our proposed ESP method with their approach in the wildlife monitoring task.

Authors of~\cite{vand} designed Multi-Agent MDP Homomorphic Networks (Equivariant MPN),  which is built upon the standard MPN. Similarly, we employed the ESP framework on the PPO algorithm used in Equivariant MPN and denoted as MPN-ESP. As depicted in Figure \ref{6b}, the proposed ESP exhibits comparable performance to that of using a Multi-Agent MDP Homomorphic Network. We conclude that our approach demonstrates similar results as network structure design methods, but it is simpler to implement and more easily adaptable to different tasks.

\section{Demonstration on robots}

As shown in Figure \ref{physical}, the trained policies were deployed on the E-puck, which is a small, lightweight, open-source robot platform. The formation change task presented in Section 6.1 was used in this experiment. We followed a direct sim2real paradigm to deploy the policy network~\cite{s2rphy}. Trajectories from a successful demonstration are displayed in Figure \ref{6c}. Different colors denote the trajectories of different robots, and color transparency indicates the temporal state. By incorporating our ESP approach into the MAPPO algorithm, the agents are able to complete tasks with fewer risky states. Risky states are defined as those in which the distance between agents is less than 5 centimeters, and the rate of risky states is the proportion of risky states to all states. The rate of risky states for ESP-MAPPO is 2.2\%, while the rate for MAPPO is 5.8\%. Additional details and videos are provided in Sec.6 of the Supplementary Material.

\begin{figure}[htbp]
     \centering
     \begin{subfigure}[b]{0.155\textwidth}
         \centering
         \includegraphics[width=\textwidth]{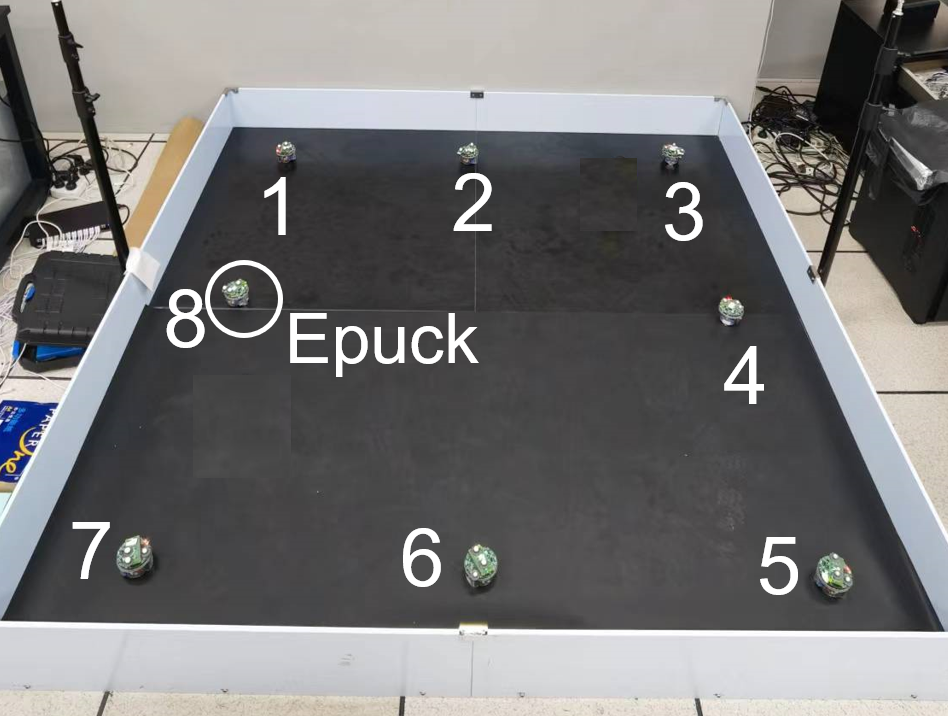}
         \caption{Start points}
         \label{fig:y equals x}
     \end{subfigure}
     \hfill
     \begin{subfigure}[b]{0.155\textwidth}
         \centering
         \includegraphics[width=\textwidth]{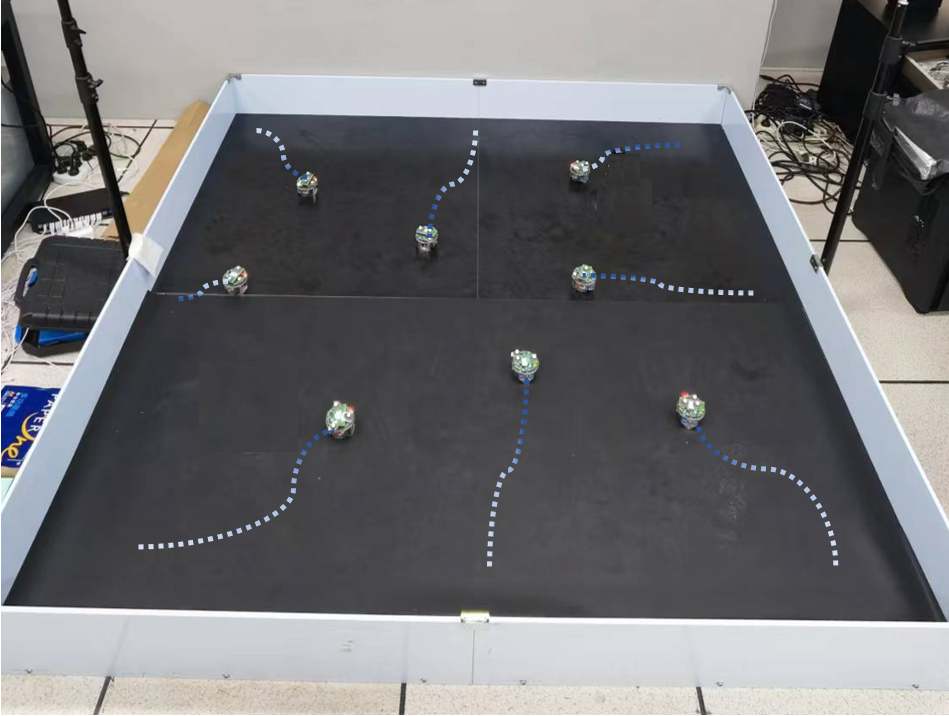}
         \caption{Trajectories}
         \label{fig:three sin x}
     \end{subfigure}
     \hfill
     \begin{subfigure}[b]{0.155\textwidth}
         \centering
         \includegraphics[width=\textwidth]{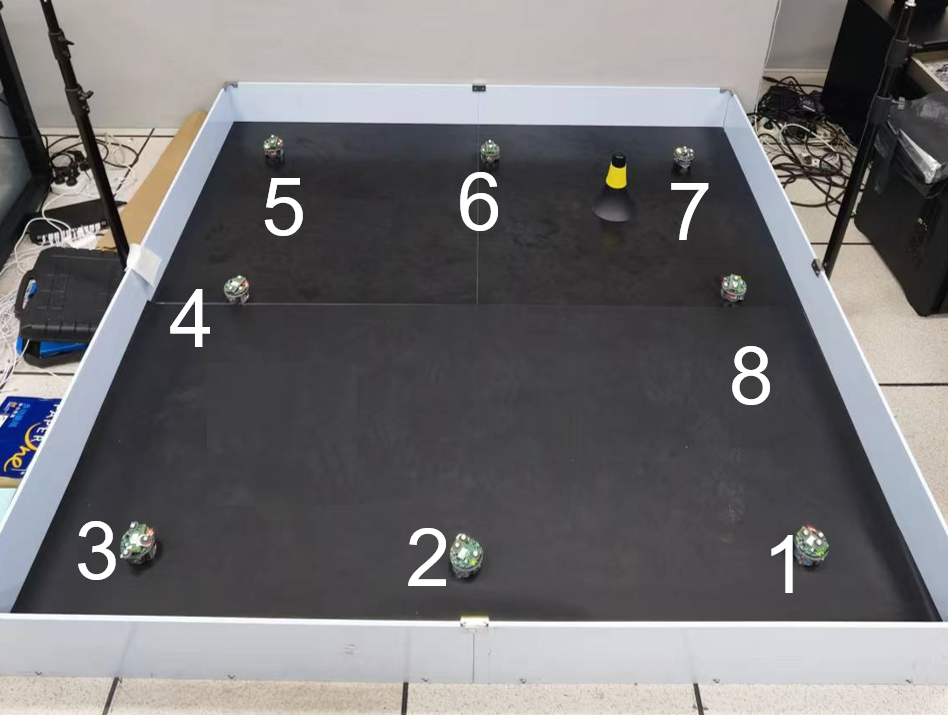}
         \caption{End points}
         \label{fig:five over x}
     \end{subfigure}
        \caption{Real-world formation change on a swarm of robots. The robots successfully switched their positions to the antipodal points by achieving collision avoidance.}
        \label{fig:three graphs}
\end{figure}\label{physical}

% 画一个真实轨迹的图
% 画一个加和不加 到目标点距离下降的图

\section{Conclusions}
In this paper, we proposed a general framework named ESP that utilizes prior knowledge to increase sample efficiency in multi-agent reinforcement learning. Our framework includes two key components: symmetry augmentation and symmetry consistency loss. By performing symmetric transformations on state-action pairs, our method generates additional data, which is stored in the replay buffer for training. The symmetry consistency loss serves as an auxiliary module to incorporate symmetry representation into the RL agent. Extensive empirical experiments on various multi-agent tasks demonstrate that our ESP framework can effectively enhance the performance of existing MARL algorithms. Moreover, our approach outperforms network structure design methods while being more straightforward to implement.

However, our approach is currently limited to scenarios where the presence of symmetry is known. Future work can extend our approach to more complex real-world tasks with unknown prior knowledge by incorporating automatic symmetry discovery functionality. Additionally, considering problems that only have local symmetry or break global symmetry to a certain degree could be a promising direction. Moreover, the integration of symmetry principles in the enhancement of robustness within multi-agent systems emerges as a noteworthy direction of research\cite{guo2022towards}. Overall, our work sheds light on the application of reinforcement learning in complex physical environments and provides a useful framework for incorporating prior knowledge in multi-agent settings.

\ack This work is supported by the National Key Research and
Development Project of China (No. 2022ZD0117801).

\bibliography{ecai}

\ifarXiv
    \foreach \x in {1,...,\numbersupplementpages}
    {
        \ifodd\x
            \includepdf[pages={\x},noautoscale=true,pagecommand={},offset=1.6cm 0cm]{\supplementfilename}
        \else
            \includepdf[pages={\x},noautoscale=true,pagecommand={},offset=-1.6cm 0cm]{\supplementfilename}
        \fi
    }
\fi

\end{document}